\PassOptionsToPackage{unicode}{hyperref}
\PassOptionsToPackage{hyphens}{url}
\documentclass[
  a4paper,
]{article}
\usepackage{amsmath,amssymb}
\usepackage{lmodern}
\usepackage{iftex}
\ifPDFTeX
  \usepackage[T1]{fontenc}
  \usepackage[utf8]{inputenc}
  \usepackage{textcomp} 
\else 
  \usepackage{unicode-math}
  \defaultfontfeatures{Scale=MatchLowercase}
  \defaultfontfeatures[\rmfamily]{Ligatures=TeX,Scale=1}
\fi
\IfFileExists{upquote.sty}{\usepackage{upquote}}{}
\IfFileExists{microtype.sty}{
  \usepackage[]{microtype}
  \UseMicrotypeSet[protrusion]{basicmath} 
}{}
\makeatletter
\@ifundefined{KOMAClassName}{
  \IfFileExists{parskip.sty}{%
    \usepackage{parskip}
  }{
    \setlength{\parindent}{0pt}
    \setlength{\parskip}{6pt plus 2pt minus 1pt}}
}{
  \KOMAoptions{parskip=half}}
\makeatother
\usepackage{xcolor}
\IfFileExists{xurl.sty}{\usepackage{xurl}}{} 
\IfFileExists{bookmark.sty}{\usepackage{bookmark}}{\usepackage{hyperref}}
\hypersetup{
  pdftitle={Bit-Twiddling Hacks for Gamma Matrices},
  pdfauthor={Thomas Fischbacher},
  hidelinks,
  pdfcreator={LaTeX via pandoc}}
\usepackage[top=30mm,left=30mm,right=30mm]{geometry}
\usepackage{color}
\usepackage{fancyvrb}

\DefineVerbatimEnvironment{Highlighting}{Verbatim}{commandchars=\\\{\}}
\newenvironment{Shaded}{}{}

\newcommand{\BuiltInTok}[1]{\textcolor[rgb]{0.00,0.50,0.00}{#1}}

\newcommand{\CommentTok}[1]{\textcolor[rgb]{0.38,0.63,0.69}{\textit{#1}}}

\newcommand{\ControlFlowTok}[1]{\textcolor[rgb]{0.00,0.44,0.13}{\textbf{#1}}}

\newcommand{\DecValTok}[1]{\textcolor[rgb]{0.25,0.63,0.44}{#1}}

\newcommand{\ImportTok}[1]{\textcolor[rgb]{0.00,0.50,0.00}{\textbf{#1}}}

\newcommand{\KeywordTok}[1]{\textcolor[rgb]{0.00,0.44,0.13}{\textbf{#1}}}
\newcommand{\NormalTok}[1]{#1}
\newcommand{\OperatorTok}[1]{\textcolor[rgb]{0.40,0.40,0.40}{#1}}
\newcommand{\OtherTok}[1]{\textcolor[rgb]{0.00,0.44,0.13}{#1}}
\newcommand{\PreprocessorTok}[1]{\textcolor[rgb]{0.74,0.48,0.00}{#1}}

\newcommand{\SpecialCharTok}[1]{\textcolor[rgb]{0.25,0.44,0.63}{#1}}
\newcommand{\SpecialStringTok}[1]{\textcolor[rgb]{0.73,0.40,0.53}{#1}}
\newcommand{\StringTok}[1]{\textcolor[rgb]{0.25,0.44,0.63}{#1}}

\newlength{\cslhangindent}
\newlength{\csllabelwidth}
\newlength{\cslentryspacingunit} 
\newenvironment{CSLReferences}[2] 
 {
  \setlength{\parindent}{0pt}
  \ifodd #1
  \let\oldpar\par
  \def\par{\hangindent=\cslhangindent\oldpar}
  \fi
  \setlength{\parskip}{#2\cslentryspacingunit}
 }%
 {}
\usepackage{calc}

\newcommand{\CSLLeftMargin}[1]{\parbox[t]{\csllabelwidth}{#1}}
\newcommand{\CSLRightInline}[1]{\parbox[t]{\linewidth - \csllabelwidth}{#1}\break}

\ifLuaTeX
  \usepackage{selnolig}  
\fi

\title{Bit-Twiddling Hacks for Gamma Matrices}
\author{Thomas Fischbacher\footnote{Google Research, Brandschenkestrasse
  110, 8002 Zürich, Switzerland - \texttt{tfish@google.com}}}
\date{}

\begin{document}
\maketitle
\begin{abstract}
\noindent For some research questions that involve Spin(p, q)
representation theory, using symbolic algebra based techniques might be
an attractive option for simplifying and manipulating expressions. Yet,
for some such problems, especially as they arise in the study of various
limits of M-theory (such as dimensional reductions), the complexity of
the resulting expressions can become computationally challenging when
using popular symbolic algebra packages in a straightforward manner.

\vskip1em\noindent This work discusses some general properties of Gamma
matrices that are computationally useful, down to the level of what one
would call ``bit-twiddling hacks'' in computer science. It is presented
in a self-contained way that should be accessible to both physicists and
computer scientists. Code is available alongside the TeX source of the
preprint version of this article on arXiv.
\end{abstract}

\hypertarget{introduction}{%
\subsection{Introduction}\label{introduction}}

The motivation for this work originated from the problem to simplify
quartic fermionic terms that arise in a Hamiltonian obtained by
dimensional reduction of 1+10-dimensional supergravity to 1+0
dimensions. Such dimensional reductions are of scientific interest as
one in general observes \emph{both} regular general relativity as well
as its higher-dimensional (possibly supersymmetric) cousins to have
extra hidden symmetries which to this date are still not fully
understood and a topic of active research. Often, extra symmetry
structure that is not obvious in some given gravitational theory then
becomes partially visible under some form of dimensional reduction.

A famous early such example arises when dimensionally reducing
Einstein's General Relativity by considering axially symmetric and
stationary solutions. Here, it can be shown that solutions of the
resulting two-dimensional theory transform under two different \(SL(2)\)
symmetries, the Matzner-Misner group {[}1{]} and the Ehlers group
{[}2{]}, which do \emph{not} commute with one another and give rise to
the infinite-dimensional Geroch group {[}3{]} (for a pedagogical
introduction -- see e.g. {[}4{]}). While we do not yet have a complete
understanding, similar phenomena are observed to occur in other
dimensionality-reducing limits of models that contain General
Relativity, such as in the Belinskii-Khalatnikov-Lifshitz (BKL) limit
{[}5{]}, and generalizations to higher dimensions and models of general
relativity with extra fields and symmetries {[}6{]}.

As major physical progress typically comes from a deeper understanding
of an underlying principle -- often a symmetry principle --, the
non-obvious symmetries of gravitational models certainly look like
interesting objects to study in the effort to construct a viable theory
of quantum gravity. Here, it is entirely plausible that even the study
of models that in themselves have little physical relevance may lead to
the development of mathematical techniques and notions that are then
more useful phenomenologically. If, for some reason, some of the happy
accidents that led the theoretical physics community to develop the
notion of nonabelian gauge theories had not happened, they readily would
also have ultimately been found in the study of dimensional reductions
of supergravity (where they arise rather naturally), offering a second
pathway for discovery.

The effort to study hidden symmetries of M-Theory then also provides the
context in which the calculation underlying this work became relevant:
can one see traces of the conjectured infinite-dimensional symmetries of
M-Theory by compactifying 11-dimensional supergravity (a low-energy
truncation of M-Theory) down to a model that only retains the time
direction -- following the general direction of prior analyses in e.g.
{[}7{]} and {[}8{]}?

Given that the present work will have some content that may be
unfamiliar to many members of the theoretical physics community, but
will be familiar to readers with a strong background in numerical
methods and computation, it is expected that readers in physics trying
to utilize the concepts and ideas described herein may want to talk to
computer scientists who in turn might find the underlying mathematics
unfamiliar. Since the problem is conceptually nevertheless rather basic
and essentially only a question of ``school level arithmetics on
steroids'', we discuss relevant background and context to readers
unfamiliar with the physical aspects in the
\protect\hyperlink{appendix}{appendix}.

\hypertarget{generic-form-of-the-hamiltonian}{%
\subsection{Generic form of the
Hamiltonian}\label{generic-form-of-the-hamiltonian}}

Generally speaking, when encountering a technically challenging
computation, it often makes sense to invest effort into contemplating
the overall structure of the problem at the ``what terms can come up for
what reason'' level, and sometimes, such analysis allows for dramatic
reduction of effort. Nevertheless, being able to quickly and
conveniently generate and collect hundreds of millions of terms
polynomial in fermionic variables in a Hamiltonian in mere minutes on a
conventional laptop is certainly an appealing capability to have
irregardless. This is the focus of the present article.

We will be generally concerned with collecting all the terms that are
quartic in four fermionic variables in a Hamiltonian, using an explicit
basis for fermionic variables and gamma matrices. These contributions
will overall come in forms such as (showing all indices explicitly):

\begin{equation}
\begin{array}{l}
\bar\Psi_{\hat RA}\Gamma^{\hat R\hat S\hat I\hat J\hat K\hat L}_{AB}\Psi_{\hat SB} \bar\Psi_{\hat TC}\Gamma^{\hat T\hat U\hat I\hat J\hat K\hat L}_{CD}\Psi_{\hat UD},\\
\bar\Phi_{A}\Gamma^{\hat 0\hat I}_{AB}\Psi_{\hat JB} \bar\Phi_{C}\Gamma^{\hat 0\hat I}_{CD}\Psi_{\hat JD},\\
\bar\Psi_{\hat IA}\Gamma^{\hat J}_{AB}\Psi_{\hat KB} \bar\Psi_{\hat IC}\Gamma^{\hat J}_{CD}\Psi_{\hat KD},\;\mbox{etc.}
\end{array}
\label{eq:terms}\end{equation}

Here, we started from the (Gamma-)trace-free vector-spinor in 11
spacetime dimensions, \(\Psi_{IA}\) i.e.~the gravitino, carrying an
11-dimensional spacetime index \(I\) as well as a fundamental
\(Spin(1, 10)\)-index \(A\). For this particular spin group, we can pick
a convenient basis in which the irreducible spinor representation (a
Majorana spinor) is all real and 32-dimensional.

Post dimensional reduction to only time, the hatted indices are
\(Spin(10)\) vector indices in internal space, and the \(\hat 0\)-index
indicates a \(Spin(10)\) scalar from the timelike component of the
\(Spin(1,10)\) vector, using a basis in which the 11-dimensional
spacetime metric is diagonal \((-1, +1, \ldots, +1)\). Since we are
using an orthonormal coordinate basis for 10-dimensional space, we make
no distinction between ``upper'' and ``lower'' \(Spin(10)\)-indices.

As usual, we have \(\bar\Phi_{B} = \Phi_A \cdot i\Gamma^0_{AB}\) and
\(\bar\Psi_{\hat I A} = \Psi_{\hat I_A}\cdot i\Gamma^0_{AB}\). We
consider the \((11-1)\times 32\) entries of \(\Psi_{\hat IA}\) to be
fundamental fermionic variables -- 320 in total. The \(\Phi_A\) are then
defined as linear functions of these that have been identified as useful
in earlier work, with each of
\(\Phi_{A=0}, \Phi_{A=1}, \ldots, \Phi_{A=31}\) being a sum of 10 of the
fundamental \(\Psi_{aB}\), each with a coefficient of \(\pm 1\). Details
beyond this basic fact are here not relevant for the computational
aspects of the problem in focus here, but can be found in the discussion
of Eq. (3.9) and (3.10) in {[}9{]} -- and also Eq. (2.36) of {[}7{]} for
a related analysis of a compactification of \(D=4\) supergravity and Eq.
(2.22) of {[}8{]} for a compactification of \(D=5\) supergravity.

Overall, the fermionic sector is of special interest for studying hidden
symmetries of general relativity and its extensions due to evidence
{[}10{]}, {[}11{]} that for D=11 supergravity, the fermionic fields
might be part of an infinite-dimensional representation of the
involutory subalgebra \(K(\mathfrak{e}_{10})\) of \(\mathfrak{e}_{11}\).

We proceed to discuss various aspects of the problem related to
computational efficiency one-by-one.

\hypertarget{d11-gamma-matrices}{%
\subsubsection{D=11 Gamma Matrices}\label{d11-gamma-matrices}}

In alignment with Eq. (3.111) of {[}12{]}, we use the following explicit
basis for the \(Spin(1, 10)\) Gamma matrices \(\Gamma^I_{AB}\):

\begin{equation}
\begin{array}{c}
  \begin{array}{cc}
  \sigma_0:=\left(\begin{array}{rr}+1&0\\0&+1\end{array}\right)&
  \sigma_x:=\left(\begin{array}{rr}0&+1\\+1&0\end{array}\right)\\
  \sigma_y:=\left(\begin{array}{rr}0&-i\\+i&0\end{array}\right)&
  \sigma_z:=\left(\begin{array}{rr}+1&0\\0&-1\end{array}\right)\\
  \end{array}\\
\\
K(P,Q,R,S,T)_{AB}:=P_{ip}Q_{jq}R_{kr}S_{\ell s}T_{m t}\delta_{A,2^4i+2^3j+2^2k+2^1\ell+2^0m}\delta_{B,2^4p+3^3q+2^2r+2^1s+2^0t}\\
\\
\begin{array}{lclclcl}
\Gamma^{I=0}_{AB}&=&K(i\sigma_y, \sigma_y, \sigma_y, \sigma_y, \sigma_y)_{AB},&\quad&
\Gamma^{I=1}_{AB}&=&K(\sigma_x, \sigma_y, \sigma_y, \sigma_y, \sigma_y)_{AB},\\
\Gamma^{I=2}_{AB}&=&K(\sigma_z, \sigma_y, \sigma_y, \sigma_y, \sigma_y)_{AB},&\quad&
\Gamma^{I=3}_{AB}&=&K(\sigma_0, \sigma_x, \sigma_0, \sigma_0, \sigma_0)_{AB},\\
\Gamma^{I=4}_{AB}&=&K(\sigma_0, \sigma_z, \sigma_0, \sigma_0, \sigma_0)_{AB},&\quad&
\Gamma^{I=5}_{AB}&=&K(\sigma_0, \sigma_y, \sigma_y, \sigma_x, \sigma_0)_{AB},\\
\Gamma^{I=6}_{AB}&=&K(\sigma_0, \sigma_y, \sigma_y, \sigma_z, \sigma_0)_{AB},&\quad&
\Gamma^{I=7}_{AB}&=&K(\sigma_0, \sigma_y, \sigma_x, \sigma_0, \sigma_y)_{AB},\\
\Gamma^{I=8}_{AB}&=&K(\sigma_0, \sigma_y, \sigma_z, \sigma_0, \sigma_y)_{AB},&\quad&
\Gamma^{I=9}_{AB}&=&K(\sigma_0, \sigma_y, \sigma_0, \sigma_y, \sigma_x)_{AB},\\
\Gamma^{I=10}_{AB}&=&K(\sigma_0, \sigma_y, \sigma_0, \sigma_y, \sigma_z)_{AB}&
&&\\
\end{array}
\end{array}
\label{eq:gamma11}\end{equation}

Generically, for any kind of computation that can in the end be reduced
to trace-over-Gamma-matrices form, it is usually advantageous to perform
term manipulations making use of the Clifford algebra to the largest
extent possible, and never going down to an explicit matrix
representation. Here, our focus however is on what to do if this
approach is either not an option, or inconvenient, or one wants to check
a result via an explicit matrix calculation.

Let us make our first computationally relevant observation. Fixing the
vector-index value, each such ``Gamma-matrix'' is the tensor (Kronecker)
product of five \(2\times 2\) matrices that each have one nonzero entry
per row and also per column. This then means that this property carries
over to the tensor product. Furthermore, since each nonzero entry in the
\(2\times 2\) matrices entering the tensor product has magnitude 1, this
also holds for each such tensor-product matrix. Finally, we trivially
observe that in the above definition, there always is an even number of
purely-imaginary factors in the tensor product, and hence our
\(Spin(1,10)\) Gamma matrices are indeed all real.

Given that for each value of the 11-dimensional spacetime vector index
\(I\), \(\Gamma^I_{AB}\) can be interpreted as a matrix with entries
only from the set \(\{-1,0,+1\}\), and exactly one nonzero entry per row
and also per column, we conclude that right-contraction (and
correspondingly then also left-contraction) with a spinor merely
shuffles the coordinates of the spinor in a prescribed way, and
subsequently flips some signs. We hence observe that, if we are only
interested in right-contracting Gammas and spinors, we can represent
each such Gamma-matrix (with index ranges in \(0\ldots 31\)) as a very
compact in-memory object that fits into a
\texttt{{[}2,\ 32{]}-int8-array}, and hence the set of 11 as a
\texttt{{[}11,\ 2,\ 32{]}-uint8-array} (a mere 704 bytes of memory, plus
some bookkeeping overhead), which we will call \texttt{g11}, indexed by
\texttt{{[}spacetime\_index,\ permutation\_or\_signs,\ spinor\_index{]}},
where (using NumPy notation) \texttt{{[}k,\ 0,\ :{]}} represents the
index-permutation to be performed on a spinor when right-contracting
with \(\Gamma^k_{AB}\), and \texttt{{[}k,\ 1,\ :{]}} is a vector of 32
signs to multiply with element-by-element afterwards. This piece of
NumPy code then shows how to apply a Gamma-matrix to a spinor,
respectively chain-multiply two Gamma matrices, using this compact
representation.

\begin{Shaded}
\begin{Highlighting}[]

\KeywordTok{def}\NormalTok{ gamma\_spinor(gamma, spinor):}
  \CommentTok{"""Computes (Gamma\^{}I)\_AB phi\_B."""}
  \CommentTok{\# Ensure we are processing numpy.ndarray objects.}
\NormalTok{  gamma }\OperatorTok{=}\NormalTok{ numpy.asarray(gamma, dtype}\OperatorTok{=}\NormalTok{numpy.int8)}
\NormalTok{  spinor }\OperatorTok{=}\NormalTok{ numpy.asarray(spinor)}
  \CommentTok{\# Indexing a numpy{-}array with an int{-}array performs}
  \CommentTok{\# a \textasciigrave{}gather\textasciigrave{} operation.}
\NormalTok{  gamma\_indices, gamma\_signs }\OperatorTok{=}\NormalTok{ gamma}
  \ControlFlowTok{return}\NormalTok{ gamma\_signs }\OperatorTok{*}\NormalTok{ spinor[gamma\_indices]}

\KeywordTok{def}\NormalTok{ multiply\_gammas(gamma\_left, gamma\_right):}
  \CommentTok{"""Computes (Gamma\^{}I)\_AB (Gamma\^{}J)\_BC from Gamma\^{}I, Gamma\^{}J."""}
\NormalTok{  gamma\_left }\OperatorTok{=}\NormalTok{ numpy.asarray(gamma\_left, dtype}\OperatorTok{=}\NormalTok{numpy.int8)}
\NormalTok{  gamma\_right }\OperatorTok{=}\NormalTok{ numpy.asarray(gamma\_right, dtype}\OperatorTok{=}\NormalTok{numpy.int8)}
\NormalTok{  gamma\_le\_indices, gamma\_le\_signs }\OperatorTok{=}\NormalTok{ gamma\_left}
\NormalTok{  gamma\_ri\_indices, gamma\_ri\_signs }\OperatorTok{=}\NormalTok{ gamma\_right}
\NormalTok{  result }\OperatorTok{=}\NormalTok{ numpy.zeros\_like(gamma\_left)}
\NormalTok{  result[}\DecValTok{0}\NormalTok{, :] }\OperatorTok{=}\NormalTok{ gamma\_ri\_indices[gamma\_le\_indices]}
\NormalTok{  result[}\DecValTok{1}\NormalTok{, :] }\OperatorTok{=}\NormalTok{ (gamma\_le\_factors }\OperatorTok{*}
\NormalTok{                  gamma\_ri\_factors[gamma\_le\_indices])}
  \ControlFlowTok{return}\NormalTok{ result}
\end{Highlighting}
\end{Shaded}

Clearly, the product of a permutation-with-subsequent-sign-flips with
another one also can be expressed in this form -- and the above code
exploits this. This brings us to our second (very simple)
computationally relevant observation: considering the general Clifford
algebra properties of Gamma matrices,
\(\Gamma^i_{AB}\Gamma^j_{BC}+\Gamma^j_{AB}\Gamma^i_{BC}=2\eta^{ij}\),
with \(\eta^{ij}\) the Minkowski metric, if our conventions put
\(\eta^{ij}\) into diagonal form (in particular, Sylvester normal form),
we obviously have
\(\Gamma^i_{AB}\Gamma^j_{BC}=-\Gamma^j_{AB}\Gamma^i_{BC}\) wherever
\(i\neq j\). So, in order to define the antisymmetrized
\(\Gamma^{ij}_{AC}:=\frac{1}{2}\left(\Gamma^i_{AB}\Gamma^j_{BC}-\Gamma^j_{AB}\Gamma^i_{BC}\right)\),
\emph{we do not actually have to antisymmetrize} and can simply take,
for \(i\neq j\), \(\Gamma^{ij}_{AC}:=\Gamma^i_{AB}\Gamma^j_{BC}\).

Likewise, since we can always rearrange chain-products of three (or
more) Gamma-matrices with mutually different spacetime-vector-indices
merely by sign-flipping for every adjacent-index-swapping, \emph{none}
of the multi-index Gamma matrices actually require us to do any
antisymmetrizations for their computation -- we can simply take the
product of the individual \(\Gamma\)s, which, like them, will ultimately
be representable as a pair of compact signed-integer vectors, one
representing a permutation, the other one the subsequent sign-flips.

In total, there are
\(\binom{D}{0}+\binom{D}{1}+\ldots+\binom{D}{D}=2^D\) such
0-spacetime-index, 1-spacetime-index, etc. Gamma-matrices in \(D\)
spacetime dimensions. So, for \(D=11\), all Gammas easily fit into
\(128\,{\rm kiB}\), and even \(64\,{\rm kiB}\) if we squeeze the 1-bit
sign into the same octet as the 5-bit permutation-index. If one is going
for such a compact representation, the only remaining problem is then
about efficiently ranking the totality of all combinations of
\(k\)-of-\(D\) different indices. This is still beyond the L1 data cache
size for most modern microprocessors, so it may be advantageous in
applications that require both heavy use of Gamma matrices and at the
same time very aggressive optimization to order calculations into blocks
in terms of the number of indices on the Gammas they use. (This often is
a structure that arises naturally).

In cases where spinors are not simply Majorana, we have to be more
careful: for Majorana-Weyl spinors, the left- and right-index are
spinorial/co-spinorial, and we need to keep track of doing contractions
on same-kind indices when forming higher products, as well as of whether
both matrix indices are spinor/spinor, spinor/co-spinor, or
co-spinor/co-spinor. For (complex) Dirac Gamma matrices (but perhaps
then already for general Gamma matrices), it is advantageous to replace
the sign-factor with an additive count of extra factors \(i\) (where
modulo-4 is automatic on, for example, unsigned 8-bit counts).
Alternatively, if one can find a way to efficiently implement this on
some specific hardware, D. Knuth's base-\(2i\) number system {[}13{]}
might possibly be an attractive option here.

\hypertarget{representing-symbolic-expressions}{%
\subsubsection{Representing Symbolic
Expressions}\label{representing-symbolic-expressions}}

Computational processing of symbolic expressions generally requires
memory cells to hold symbol-references, both for identity-checking and
also indirect referencing (through the symbol definition) of further
definitions related to the corresponding symbol. Working with such
data-indirections is generally computationally costly, and so the
question needs to be evaluated whether symbolic processing of data is
actually competitive against a \emph{de-symbolizing} approach which
replaces such symbols with numerical data in an un-ambiguous way.

The underlying idea here is that numerics can be executed at very high
speed on modern microprocessors. This trade-off generally depends on the
problem, but for the case at hand, we observe that there are
\(\binom{320}{4}=428\,761\,520\) different quartic combinations of
fundamental fermionic factors. If symbolic processing represented each
combination of factors as an in-memory vector of four symbol-references
plus a non-integer coefficient, then each such symbol-reference would
worst-case require one pointer (on modern systems 8 octets (i.e.~8-bit
bytes)), but in a more carefully designed system might get away with a
reference to a symbol-table (one 8-octet pointer) and a smaller index
into that table per symbol. In principle, it would be possible to use a
compact memory representation that puts a floating point coefficient of
a term into the bookkeeping information that also stores the number of
factors, but for a generic symbolic algebra system, this might not be
available. Since the overarching expression will then also want to
reference that vector as part of a larger expression, the in-memory
representation will need another pointer per such product. This then
means that each of the worst-case about \(430\) million terms would
require between about 16 and 60 octets of memory (so in total about 6-24
GiB), which would be processed with typically irregular memory access
patterns. In our case, if we proceeded one-by-one by
spacetime-index-allocation on the Gamma matrices, we would still for
every such summand have to work out two factors, each a fermionic
bilinear expression that consists of either \(32\times32\) or
\(10\times32\times32\) summands, depending on whether there is a
\(\bar\Phi\) to the left or not, and then multiply these. Efficient
multiplication is in itself an interesting algorithmic problem (see e.g.
{[}14{]}), but the relevant algorithms are generally not readily
available in a form that would facilitate their adaptation to tailored
data processing. So, for practical purposes, somewhat efficient
implementations of such symbolic techniques are available through
(mostly commercial) stock symbolic algebra packages, but require major
coding effort to tailor them to a specific problem more closely.

Borrowing from microprocessor design terminology, this means we should
favor a ``speed demon'' over a ``brainiac'' approach. i.e.~optimizing
for a compact in-memory representation plus as-fast-as-possible explicit
iteration through all index-combinations. A ``brainiac'' strategy would
here likely employ a system like FORM {[}15{]}, or perhaps use an
equivalent approach based on some MapReduce {[}16{]} like Big Data
framework to perform efficient manipulation of expressions still at the
fully symbolic level -- but in a problem-specialized way.

Performing a de-symbolized calculation that does away with all
symbol-tags to identify factors, and rather discriminates contributions
by-index, and observing that for most problems of this nature,
single-precision floating point coefficients should work well, we can
the calculation using only about 1.6 GiB of memory for collecting
coefficients. The price to pay is that per vector-index combination on
the Gamma matrices, we would have to do a nested loop over
\(32\times32\) times either 1, or 10, or 100 contributions. Given that
modern microprocessors are fast enough to process 100 million numerical
operations in mere milliseconds, this option looks promising especially
from a memory-usage perspective. It makes sense to implement this
iteration in a fast, compiled, widely understood language such as C++,
but use Python for the higher-level orchestration of the calculation,
mostly due to the rather useful multi-index array processing semantics
of \texttt{NumPy} {[}17{]}.

The main idea behind de-symbolizing Gamma-matrix calculations is that
for both the first and second fermionic-bilinear factor in a generic
term
\((\psi_{\ldots}\Gamma^{\ldots}\psi_{\ldots})_1(\psi_{\ldots}\Gamma^{\ldots}\psi_{\ldots})_2\)
of any four-fermion term in the Hamiltonian, if we know all the
spatial-\(Spin(10)\)-vector indices, the index on the right-side
vector-spinor selects a 32-coefficient spinor, to which we then have to
apply some particular Gamma matrix. Here, we can afford memorizing how
each of the at most 2048 0-index, 1-index, etc. Gamma-matrices permute
and sign-flip each of the \(10\times 32\) vector-spinors \(\Psi_{IA}\),
since this needs no more than 0.7 million memory-cells, where each cell
is large enough to store the identity of one of the fundamental
fermionic variables and its sign, which with careful coding we can do in
two octets (explained below).

Conveniently, having looked up a particular
\(\Gamma^{I_0I_1I_2\ldots }_{AB}\Psi^{I_m}_B\) in this table by key
\((I_0I_1I_2\ldots)\), which then yields a table indexed by
\((I_m, A)\), picking a sub-table via NumPy slicing will not have to
copy any data, but will instead only add extra bookkeeping information
that there is one more numerical object accessing the already-present
in-memory data, which provides merely another view on a sub-range of
that data (but, perhaps unusually for the notion of a memory-view,
allowing mutation through this view). For the left hand side, we have to
look up either one of the \(\bar\Psi^a_A\) by vector-index, which yields
a spinor holding 32 signed fundamental anticommuting variables, or
process a \(\bar\Phi\), which we represent as a \([10,32]\)-array with
instruction to iterate over the range-10 number-of-summand index
producing quartic terms.

A useful approach to representing a size-32 vector of signed fundamental
anticommuting quantities, of which there are 320 different ones, is to
realize that applying any Gamma-matrix shuffles the vector and then
performs sign-flips, so we can keep track of these fundamental
quantities by simply representing each with a different nonzero integer.
Since we want sign-flips to be represented as sign-flips, zero must be
excluded. The most straightforward such encoding scheme would use the
integer +1 to represent the lexicographically-smallest of the 320
fundamental anticommuting variables, -1 to represent its negative, +2 to
represent the lexicographically second fundamental anticommuting
variable, -2 to represent its negative, etc. Since there are 320 such
variables in total, we cannot represent the 640 different cases in one
octet, and instead go with cells holding signed 16-bit integers (2
octets).

\hypertarget{normal-ordering-and-contribution-generation}{%
\subsubsection{Normal Ordering and
Contribution-Generation}\label{normal-ordering-and-contribution-generation}}

Per allocation of vector-indices (of which for any given term in the
Hamiltonian, there typically are only some thousands, perhaps tens of
thousands), we can identify a collection of four arrays for which we
have to form products along the spinorial axes of the first two, and
independently the second two, and perhaps also iterate over a range-10
summand-index for each of the left-side-of-a-bilinear factors that come
as a \(\bar\Phi_A\).

Naturally, such a task can be performed without requiring any dynamic
memory management. As such, The most straightforward option here for
making this part of the computation fast that requires only a modicum of
framework-level technology is to use a compiled Python module built
directly on the Python/NumPy foreign function interface.

While the basic task is simple enough for C++ to provide no relevant
advantage over C here, apart from minor conveniences, one would in
general also want to support linear substitutions on fundamental
fermionic variables which, if done in-situ, are somewhat easier to
implement in C++ by using a C++11 lambda expression to implement a
looping primitive for variable-expansion to which the loop-body is then
provided as a function, reducing substantial code duplication at the
(moderate) expense of having to handle
\texttt{std::function\textless{}\textgreater{}} objects. This approach
parallels {[}18{]}, where OCaml has been picked as a fast compiled
language both due to the absence of closures pre-C++11 and the need to
retain some symbolic processing.

ven though the task at hand is narrowly defined, a fundamental principle
one should never violate in any foreign function module is that even
providing invalid data, such as wrong-size arrays (or objects other than
arrays), or arrays with content other than the expected one, must not
produce crashes due to (for example) bad memory references. As we only
expect tens of thousands of calls to go through the interface anyhow,
investing a bit of effort to check validity of the data on the
receiver-side is not a problem. The task is overall simple enough to not
benefit in relevant ways from higher level frameworks that build on top
of the Python C API. One important detail that users of Python's foreign
function interface who want to process NumPy data arrays from the C side
must be aware of is that the in-memory data-ordering may be unusual. In
particular, if \texttt{arr} is some 2-index numpy array, then
\texttt{arr} and \texttt{arr.T} are in general merely different
``views'' on the same in-memory data-block with swapped
multi-index-to-linear-index weights. So, when accessing data, one either
has to use C-level utilities such as \texttt{PyArray\_GETPTR2()} to have
this index-mapping handled automatically, or alternatively carefully
work with \texttt{PyArray\_STRIDES()} to obtain striding information.

On the C side, we are performing four nested loops over the data, two
being range-32 for the left-bilinear and right-bilinear scalar products
between spinors, and each of the other two being either range-1 if the
corresponding left-side-of-bilinear factor is a \(\bar\Psi_{IA}\), or
range-10 for a \(\bar\Phi_A\). This way, we fetch a sequence of four
short signed integers each representing a signed fundamental fermionic
quantity. In order to determine the overall coefficient, we need to
count the number of sign flips needed to make each integer-tag positive,
multiply with the sign of the permutation required to lexically order
the four factors, and then multiply with the overall term-coefficient.
If the four factors are not all different, then that summand contributes
nothing, and this can be detected early already on partially-filled
four-fermion combinations.

When lexically ordering four factors, the naive approach would be to use
a bubblesort strategy and count swaps. While it is
mostly-straightforward to generalize mergesort in such a way that it
counts the number of swaps, we here propose a different strategy, as an
optimization: using a recursive functional algorithm (once) to generate
explicit code with nested-comparisons only that sorts four different
items. We here do this in alignment with how this technique was employed
in {[}18{]}, where a more detailed discussion can be found in the source
code accompanying the arXiv preprint.

Having sorted a quartic product, the next question is how to accumulate
coefficients per such product. Two commonly employed methods are to
either use some hash-based data structure, or merely generate
contributions, and then shuffle/reduce via a MapReduce {[}16{]} ``Big
Data'' strategy. Hashing would require storing hash-keys for comparison,
and while these could here be rather compact, even a clever approach
that regards the 4-tuple as two 2-tuples, of which there are only
\(320\cdot319/2=51040<2^{16}\), and then enumerates these, would require
as much memory for the hash keys alone as for the coefficients. A better
option would be to determine the function that maps a 4-tuple to its
index in a lexicographic ordering of all of them and use that as an
index into the coefficient-array. The code accompanying this work
contains hash-based accumulation, mostly for testing/debugging purposes
as an easy-to-understand baseline implementation, plus also fast
tuple-ranking based implementation.

The reverse-mapping from an index to a 4-tuple is here not really
needed, since scanning the coefficient-array in parallel to going
through the enumerated lexically ordered sequence of permissible
4-tuples can easily be done as a one-off operation right at the end.

While working out the generic four-tuple ranking function is a doable
combinatoric exercise, one can alternatively employ a very simple idea
whose far-reaching consequences are discussed in the amazing book A=B
{[}19{]}: one observes that, by induction, the ranking function must be
a polynomial of degree four in the four tuple-elements, which we can
determine simply by solving a linear equation system for the
coefficients of the \(\binom{4+5-1}{4}=70\) summands
\(c_{abcde}t[0]^at[1]^bt[2]^ct[3]^d\cdot1^e\) with \(a+b+c+d+e=4\). Some
care has to be taken to not pick a degenerate set of samples, and
solving an over-determined equation system can provide a first test at
the same time as finding the solution.

If one were to actually use a hash table, such as a Python
\texttt{dict}, to collect coefficients by factors-tuple (perhaps for
comparing with the speed-optimized variant), and the latter is available
as a numpy.ndarray \texttt{factors}, a fast and convenient way to come
up with a compact hash key that uniquely identifies the index
combination is to use
\texttt{factors.astype(\textquotesingle{}\textgreater{}u2\textquotesingle{}).data.tobytes()}.
In Python, \texttt{numpy.ndarray} instances are mutable and hence
normally not hashable, and the language prevents using such objects as
hash keys. (Hashing even is not defined after setting
\texttt{factors.flags.writeable=False}, which is generally not a
recommended-to-use NumPy feature anyhow). A byte-string carrying the
same data will serve well as a very compact key that is easy to hash and
compare, and by requesting most-significant-byte first order (via
\texttt{.astype(\textquotesingle{}\textgreater{}u2\textquotesingle{})}),
we even make lexicographic ordering of keys parallel lexicographic
ordering of factors. Since our index-range exceeds 255, we have to ask
for two octets per array entry.

As is often the case in computational problems with a quantum gravity
origin, coefficients have power-of-two denominators. Conveniently, this
means that floating point arithmetics is actually \emph{exact} on such
quantities, unless one exhausts the capacity of the mantissa, which a
simple argument shows to here not be the case: while we are generating
and summing hundreds of millions of terms, every specific quartic term
can only arise specific allocations of the spacetime-vector-indices on
the \(\Psi_{aA}\) and the \(\Gamma^{\cdots}\), of which there are
(per-term), even taking different coefficients that may scale by a small
number of factors 2 into account, too few to possibly exhaust even the
23-bit mantissa of IEEE-754 binary32 single-precision floating point
numbers. (If there are concerns about this point, doing the calculation
in double-float precision merely doubles memory requirements and
increases overall run time by a small factor.)

In trickier situations, the article ``What Every Computer Scientist
Should Know About Floating-Point Arithmetic'' {[}20{]} provides useful
background on how to use floating point machinery to nevertheless do
exact calculations.

One possible further optimization opportinity for the calculation that
is not implemented in the accompanying code comes from the observation
that the calculation is mostly memory I/O bound. This is readily checked
by providing a 4-tuple ranking polynomial that makes every contribution
get accumulated on the same memory-cell, rather than spread out over a
large region. So, for exclusively power-of-two denominators, going from
binary32 float to int16 (but then with a range check, to guarantee
validity of the result) coefficients may be an easy option for cutting
memory I/O in half.

\hypertarget{vector-index-tuple-generation}{%
\subsubsection{Vector-index-tuple
generation}\label{vector-index-tuple-generation}}

We are still left with the problem to iterate over all relevant
allocations of vector-index values. As one would expect, every
spatial-\(Spin(10)\) vector-index will occur twice, and either connect a
\(\Gamma\) on one fermionic bilinear with another \(\Gamma\), or with a
\(\Psi\) on the same bilinear. Going forward, we will process
bilinear-internal indices that connect a \(\Gamma\) to an adjacent
\(\Psi\) by looping over all index-allocations for such indices
separately, and then focus on indices that connect the two bilinears.

With these sorted out, we still have to generate all vector-index
allocations for vector indices shared between the left and right
fermionic bilinear in the quartic expression.

A generic such term that we may encounter might be:

\begin{equation}
-\frac{1}{2}\langle\bar\Psi^{\hat I} \Gamma^{\hat 0\hat J} \Psi^{\hat K}\rangle
\langle\bar\Psi^{\hat R} \Gamma^{\hat R\hat S\hat 0\hat I\hat J\hat K} \Psi^{\hat S}\rangle
\label{eq:indexCombination}\end{equation}

This is conveniently transcribed to Python by making use of
generator-expressions and advanced iteration, illustrated below. The
accompanying code's \texttt{QuarticTermAccumulator} API takes a Python
iterable (which may be an iterator, such as a generator-expression) that
provides (\texttt{combinatorial\_factor},
\texttt{left\_bilinear\_indices}, \texttt{right\_bilinear\_indices})
data, where for both bilinear factors, indices are given as:
(\texttt{left\_psi\_vector\_index}, \texttt{gamma\_vector\_indices},
\texttt{right\_psi\_vector\_index}).

\begin{Shaded}
\begin{Highlighting}[]
\KeywordTok{def}\NormalTok{ add\_term\_h1(accumulator):}
\NormalTok{  r1\_11 }\OperatorTok{=} \BuiltInTok{range}\NormalTok{(}\DecValTok{1}\NormalTok{, }\DecValTok{11}\NormalTok{)}
  \ControlFlowTok{for}\NormalTok{ R, S }\KeywordTok{in}\NormalTok{ itertools.product(r1\_11, r1\_11):}
\NormalTok{    accumulator.collect(}
\NormalTok{        term\_factor}\OperatorTok{={-}}\DecValTok{1}\OperatorTok{/}\DecValTok{2}\NormalTok{,}
\NormalTok{        ci}\OperatorTok{=}\NormalTok{(}
\NormalTok{            (}\OperatorTok{+}\DecValTok{1}\NormalTok{, (I, (}\DecValTok{0}\NormalTok{, J), K), (R, (R, S, }\DecValTok{0}\NormalTok{, I, J, K), S))}
            \ControlFlowTok{for}\NormalTok{ I, J, K }\KeywordTok{in}\NormalTok{ itertools.product(r1\_11, r1\_11, r1\_11)))}
\end{Highlighting}
\end{Shaded}

Depending on what other combinatorial properties we may be able to
exploit on individual expressions, we might want to use more
sophisticated generator-expressions to only iterate over some
combinations and compensate with an overall factor, perhaps nesting one
generator-expression inside another to do post-processing or filtering.

\hypertarget{other-computational-aspects}{%
\section{Other computational
aspects}\label{other-computational-aspects}}

Given advances in computing technology, even a problem of the discussed
type is nowadays too small to warrant an in-depth discussion of options
for further performance improvement, related to techniques such as:
multicore parallelization, memory-mapping the coefficient data with
(POSIX) \texttt{mmap()} and then perhaps using (Linux)
\texttt{madvise()} to fine-tune paging (such as with
\texttt{MADV\_PAGEOUT}). One relevant aspect that can start to play an
important role for very large symbolic computations that require many
CPU-hours of effort which here needs to be addressed is that bit-flips
are here generally not benign, and for hardware not using
error-correcting (ECC) RAM, ballpark figures such as those reported in
{[}21{]} of 25-70 errors per million device-hours per megabit (which
translates to about 5-14 bit errors per GiB-day, albeit often coming
from easily-detectable ``stuck bits'') become a real concern.

In some cases, the best option to address this problem is to split the
calculation into individually-reproducible parts and looking for small
adjustments to the computation that then also provide a checksum -
unless the problem already comes with some form of such property
allowing a consistency-check. Some of these techniques have been used
widely in times when computations could not be done reliably (such as
when human computers were still employed), but nowadays have been all
but forgotten. For example, adding a row and column to each matrix such
that all rows and also all columns sum to zero, this property is
preserved under matrix multiplication.

\hypertarget{conclusion}{%
\section{Conclusion}\label{conclusion}}

We have started from a generic kind of expression that is bound to come
up in many models of General Relativity with fermionic matter and, just
paying a bit of attention to the computational structure of Gamma
matrices, have broken it down along some perhaps not often trodden path
to a calculation which is observed to take less than three minutes on a
typical not-too-modern laptop -- even to a point where a competent CS/EE
master would see how to realize most of it in custom hardware (perhaps
using a modern FPGA). This gives us another pathway to go from problem
to solution for tasks of a similar nature. The code accompanying this
work is available both alongside the TeX source of this work's preprint
on arXiv, and also on github at
\href{https://github.com/google-research/google-research/tree/master/m_theory}{\texttt{https://github.com/google-research/google-research/tree/master/m\_theory}},
in the
\href{https://github.com/google-research/google-research/tree/master/m_theory/dim1/papers/gamma_bth/}{\texttt{dim1/papers/gamma\_bth/}}
directory.

One could here make the point that in physics, we have learned in the
past that the best way to approach such calculations is to avoid the
need to handle explicit forms of Gamma matrices entirely, instead
maximally exploiting the Clifford algebra properties and also
Fierz-Pauli relations. This is indeed often the case, but still, in
situations where an explicit calculation looks attractive in principle,
it is better to be able to easily do the computation than not to -- also
perhaps to have a different computational pathway that allows an
independent check of a result.

Apart from that, the bridge-building effort to make the main concepts
and ideas accessible to computer scientists without a physics
background, with whom physicists might generally want to collaborate on
a project of this nature, may additionally have helped making some of
the rationale behind investigating supersymmetric models of gravity
accessible to a broader audience, hopefully contributing to ultimately
expanding the group of professionals who can assess the validity of a
branch of research that easily can look somewhat esoteric from the
outside.

\hypertarget{acknowledgments}{%
\section{Acknowledgments}\label{acknowledgments}}

The author would like to thank Thibault Damour and the Institut des
Hautes Études Scientifiques (IHES) for hospitality, and his managers at
Google Research Zurich, Jyrki Alakuijala and Matt Sharifi, for support
on this project. This work is dedicated to Jonathan Klein, for personal
reasons.

\hypertarget{appendix}{%
\section{Appendix: Basic Concepts}\label{appendix}}

We here briefly sketch some of the relevant concepts underlying
Supergravity at a basic level for an intended target audience that might
not naturally be familiar with these concepts, such as computer
scientists collaborating with physicists. In the spirit of building
bridges between disciplines, it is worthwhile to point out that Gamma
matrices and the Clifford algebra are not merely esoteric topics that
would only be relevant to physicists dealing with relativistic theories
involving fermions, but also have been understood more widely to allow
an elegant and convenient approach to many basic geometric problems via
Geometric Algebra / Conformal Geometric Algebra, see e.g. {[}22{]} and
{[}23{]} for an overview and introduction, and {[}24{]}, {[}25{]},
{[}26{]}, {[}27{]} for a survey of practical applications.

In a nutshell, linear algebra as it is commonly taught in post secondary
education courses mostly focuses on vectors and linear mappings, and in
applications to geometry, vectors typically describe relative location
of two points (one of which may be ``the coordinate origin''). Due to
time constraints, one would on a basic course also encounter
determinants as ``volume-forms'' and perhaps co-vectors / linear forms,
but typically not talk about oriented \(k\le N\)-dimensional volumina in
\(N\)-dimensional space. So, one basic question that is typically just
beyond what is covered in such courses is how to elegantly(!) determine
for two pairs of five-dimensional vectors whether the (oriented)
parallelogram defined by the first pair has the same orientation in
space and also area as that defined by the second pair -- and
correspondingly for parallelepipeds generated by triplets of vectors.
For applications in three-dimensional geometry, the cross-product can be
employed to (ab)use 3d vectors to describe such oriented patches,
allowing one to shoe-horn the generic discussion of ``sized and oriented
geometric objects modulo shape'' that can be traced back all the Way to
Grassmann's 1862 ``Ausdehnungslehre'' {[}28{]} into a time-constrained
course. (See {[}29{]} for a discussion of Grassmann's work).

Intuitively, one can contemplate defining a formal vector space of all
oriented quantities for \(N\)-dimensional space, where the ``vectors''
form a \(N\)-dimensional subspace, the ``oriented patches'' form a
\(N(N-1)/2\)-dimensional subspace, etc., realize that sized and oriented
\(k\)-dimensional objects (modulo shape) generate a
\(\binom{N}{k}\)-dimensional subspace of ``\(k\)-vectors'', and notice
that this makes the entire space \(\sum_k\binom{N}{k}=2^N\)-dimensional.
With this, we can see a multi-index Gamma-matrix such as
\(\Gamma^{IJK}_{AB}\), with some specific \(I,J,K\), as a projector that
can extract the \((I,J,K)\)-coordinate of a \(k\)-vector (here,
3-vector) from a generic matrix with two spinorial indices, \(F_{AB}\)
via tracing
\({\rm tr}\left(\Gamma^{IJK}\,F\right)=\Gamma^{IJK}_{AB}\,F_{BA}\).

The link between Geometric Algebra language and Gamma Matrix language is
provided by realizing that e.g.~Eq. (3) of {[}22{]}, \begin{equation}
\mathbf{x}\mathbf{y}=\mathbf{x}\cdot\mathbf{y}+\mathbf{x}\wedge\mathbf{y}
\label{eq:geometricalgebra}\end{equation} or in orthonormal coordinates
(i.e.~\(g^{ij}=\delta^{ij}\)), \begin{equation}
x^ie_i y^je_j = x^iy^jg_{ij} + x^iy^j\cdot\frac{1}{2}(e_ie_j-e_je_i),
\label{eq:geometricalgebracoords}\end{equation} would, with Gamma
matrices, equivalently be written as: \begin{equation}
x_i\Gamma^i y_j\Gamma^j = x_iy_j\Gamma^{()} + x_iy_j\Gamma^{ij},
\label{eq:geometricalgebragamma}\end{equation} with the 0-index
\(\Gamma^{()}\) being the identity matrix (the product of zero Gammas).
Why would it make sense to talk about sums of (nominally) a 0-index
\(\Gamma^{()}\) and a 2-index \(\Gamma^{ij}\)? To see this, one has to
realize that the totality of \(\Gamma^{()}\), \(\Gamma^{i}\),
\(\Gamma^{ij}\), etc. allow us to define projectors on spinor-matrices
\(M_{AB}\) that satisfy a completeness relation, \begin{equation}
\begin{array}{llcll}
\bigl(&\Gamma^{()}_{A'B'}\Gamma^{()}_{AB}&+&&\\
&\Gamma^{i}_{A'B'}\Gamma^{i}_{AB}&+&&\\
&\Gamma^{ij}_{A'B'}\Gamma^{ij}_{AB}&+&&\\
&\Gamma^{ijk}_{A'B'}\Gamma^{ijk}_{AB}&+&\ldots\bigr)M_{BA}&=M_{A'B'}.
\end{array}
\label{eq:gammacomplete}\end{equation} So, the direct sum vector space
of all scalars, vectors, bi-vectors, tri-vectors etc. is merely a
re-packaging (for even dimensions) of the information content of a
bi-(Dirac-)spinorial complex \(2^{N/2}\times 2^{N/2}=2^N\) matrix into
parcels of size \(\binom{N}{0}+\binom{N}{1}+\ldots+\binom{N}{N}=2^N\).

For physicists familiar with Gamma matrices, the insight here is that by
investing only little effort into learning the term dictionary, one can
have meaningful discussions about elegant mathematical approaches to
computer vision and robotics problems with practitioners in these
fields.

\hypertarget{general-structure-of-the-theory}{%
\subsection{General Structure of the
Theory}\label{general-structure-of-the-theory}}

Overall, we observe that there are four fundamental interactions
(``forces'') in our universe. Two of these are, at the classical level,
widely known and understood: Electromagnetism and Gravity. The modern
description of these phenomena uses the framework of gauge theory,
albeit in two different forms: For electromagnetism, the gauge boson,
the photon, is a particle with helicity-eigenstates of angular momentum
\(\pm\hbar\). For gravity, the gauge boson, the graviton, has to have
helicity-eigenstates with angular momentum \(\pm2\hbar\). The other two
forces in nature, which govern in particular phenomena related to
nuclear structure and decay, the Strong Force and the Weak Force, are
also described by gauge theories for which one uses some modification of
the framework of helicity-\(\pm\hbar\) gauge particles - as for
electrodynamics. Overall, gauge theories are seen as rather fundamental
building blocks of physical models; one would generally not regard gauge
symmetry to be ``accidental''.

In principle, one could (at least classically) imagine an
Einstein-Maxwell ``toy model world'' in which the only objects are
electrically charged black holes, and the only forces are gravity and
electromagnetism. Such a world would not have ``matter'' as we know it,
where we want to consider the defining property for ``matter'' to not be
that particles have to have mass, but that when trying to collectively
confine particles to some limited space, there is a contribution to
pressure that comes from the collection of particles resisting
compression. While adding photons to a perfectly mirrored box would
increase light pressure on the wall, if we are free how to add the next
photon, we always can add it to the energetic ground state, at a
constant increase of light pressure per photon on the walls. In
contradistinction, for fermions, the increase of pressure depends on the
number of fermions already present in the box.

Tantalizingly, we both observe that our world \emph{does} have matter,
and also that, perhaps as a mere mathematical curiosity, there appears
to be exactly one further flavour of mathematically (likely) viable
gauge theory beyond the two that give us gravitation and
electromagnetism (alongside its two weak/strong force cousins). This
third kind of gauge symmetry would come with gauge particles of helicity
\(\pm(3/2)\hbar\), which then would not be gauge bosons, but \emph{gauge
fermions}. One then finds that any mathematically consistent models of
physics involving this type of gauge symmetry would not only have to
have matter particles, but on top of that would also have to include
gravitation: these models are called Supergravities, and the new kind
gauge particle would be called a ``gravitino''. In 1975, it was realized
that a model of gauginos is mathematically compatible (and even
requires) general relativity {[}30{]} (which, indicentally, in this work
was demonstrated with at-the-time-heroic symbolic calculation with
tailored-to-the-task code -- that however was soon after rendered
unnecessary by a clever argument, {[}31{]}). Soon after, this led to the
insight that general relativity mathematically admits extensions with
more than the minimal amount of supersymmetry, where (super)symmetry
multiplets then stretch across larger groups of fundamental fields
(i.e.~particles). As the original attempt to construct the maximally
symmetric such extension ran into technical difficulties, a detour was
attempted {[}32{]} that (successfully) first tried to construct a
supersymmetric model of gravity in its maximal possible spacetime
dimension (constructed at that time entirely as a mathematical reasoning
device), and then dimensionally reduce this to four-dimensional
spacetime.

Overall, symmetries of physical models give rise to conservation laws
(via Noether's theorem). A basic way to observe this is to realize that
conservation of electrical charge and the inability to come up with an
experiment that would authoritatively define ``electric ground
potential'' go hand in hand: If we ascribe to an electric charge \(q\)
at potential \(\Phi\) the energy \(q\cdot\Phi\), then destroying one
unit of charge would require a particular energy that depends on the
value of the potential, and we could consider that electric potential
for which charge can be destroyed at zero energetic cost as being
special and singled out by a physical experiment. As such, models with a
lot of symmetries may perhaps look formidable when seeing the equations
of motion written down on paper, but in some situations would behave in
a conceptually easier-to-understand way than nominally less complicated
models, due to being constrained by a large number of conservation laws.
As such, one in particular observes scattering to often be easier to
analyze in supersymmetric variants of General Relativity than in General
Relativity itself, hence providing useful checks and benchmarks for
calculational tools and techniques {[}33{]}.

Overall, in the grand scheme of things, there are multiple reasons to
consider supersymmetry an interesting idea. For quite a few problems, it
is a useful mathematical (but really merely mathematical) tool for which
it would be as absurd to ask for an experimental proof of its existence
as it would for complex-valued voltages and currents as they are widely
used in electronics -- which are a mathematical tool in a very similar
spirit.

\hypertarget{on-the-role-of-anticommuting-numbers}{%
\subsection{On the role of ``anticommuting
numbers''}\label{on-the-role-of-anticommuting-numbers}}

There are two complementary and equivalent approaches to quantum field
theory: the operator-based approach follows the construction of
nonrelativistic quantum mechanics and emphasizes unitarity
(i.e.~``conservation-of-probability'', which any mathematically
meaningful theory must have), at the expense of making compatibility
with Special Relativity somewhat non-obvious. The basic conceptual
problem here is that if we want to talk about the probability-amplitudes
for a particle to be in some particular region of space or some other
such region, this involves an implicit concept of simultaneity --
``different regions at the same time''. While this per se does not yet
lead to a conflict with Relativity, the key question is whether physical
claims made about measurements by observers moving relative to one
another are compatible.

The alternative picture, the ``path integral formalism'' popularized in
particular by Feynman, emphasizes compatibility with Relativity, but at
the expense of making preservation-of-probability more painful to keep
track of. For a viable quantum field theory, these two pictures are
equivalent to one another despite the conceptual tension between them,
but the problem that it is difficult to have both manifest unitarity and
manifest relativistic covariance in the same description does not appear
to be a mere issue of finding a convenient language: in string theories,
one finds that the effort to reconcile the two relevant properties
introduces a subtle constraint on the dimensionality of spacetime
itself.

Overall, the ``operator formalism'' approach is mostly considered to
give rise to (in direct comparison) ``too messy'' expressions to be used
much in quantum field theory, with virtually all work being done via the
path integral approach. Some general intuition about the relation
between these approaches can be obtained by considering how one would
evolve an initial-time quantum state \(|\psi_{t_i}\rangle\) to a
final-time quantum state \(|\psi_{t_f}\rangle\) over 10 time steps, each
implemented by the application of some unitary matrix, intuitively
\(\hat U=\exp -i\hbar \hat H\Delta t\), with \(\hat H\) the Hamilton
operator of the quantum system. For a finite-dimensional quantum system
(which just about all quantum field theories are not), one would
introduce an orthonormal basis for state-space and describe a quantum
state \(|\psi\rangle\) as a linear combination of basis vectors,
\(|\psi_t\rangle=\sum_i {\mathtt a}_{t,i} |\psi^{(B)}_i\rangle\). With
respect to the same basis, \(\hat U\) would then be represented as a
complex unitary matrix \(\mathtt{U}\). In the operator formalism
approach, one would consider each time-step to be represented by a
matrix-vector multiplication, which for the coefficients means
\({\mathtt a}_{t+1, i}=\sum_k \mathtt{U}_{ik}\mathtt{a}_{t, k}\), i.e.
\(\vec {\mathtt a}_{t_f}=U(U(\cdots U(U\vec {\mathtt a}_{t_i})\dots))\).

In the path integral formalism, one would instead focus attention on all
the possible ways to obtain a contribution to \({\mathtt a}_{t+1, i}\)
by taking all possible routes through the ten \(\mathtt{U}\)-matrices,
i.e. think about this expression as if it were written in the form
\begin{equation}
{\mathtt a}_{t_f, i_{10}} = \sum_{i_0,i_1,i_2,\ldots,i_9} {\mathtt U_{i_{10}i_9} U_{i_9i_8}\cdots U_{i_1i_0} {\mathtt a}_{t_i, i_{0}}}
\label{eq:pathintegral}\end{equation} In a quantum field theory, the
individual summands can be naturally interpreted as ``classical paths''
if the quantum state space basis is aligned with classically measurable
quantities (such as particle-position), i.e.~the eigenvalues of
operators representing observables.

This approach causes a problem with path integrals involving nilpotent
operators, for which the algebraic multiplicity of eigenvalues differs
from their geometric multiplicity. These arise when describing fermions.
Without getting too technical, some intuition about how the mathematical
resolution to this problem behaves physically can be obtained by looking
at desiderata. When we are looking at a quantum system that contains two
identical-type fermions, such as two electrons, we would like to
maintain the property that if we swapped the role of two electrons, then
the quantum amplitude would change sign. While there is a deep
mathematical reason (the spin-statistics theorem) at work here that we
will not look into, what this accomplishes is that whenever two such
fermions try to go into the same quantum state, the amplitude for the
process where this happens will be zero -- one could say that it cancels
with the amplitude for the process where the two particles enter the
same state with roles-reversed. This way, two fermions cannot enter
(scatter) into the same quantum state, in alignment with the observation
that matter resists compression in different ways than radiation (in
terms of what happens to average particle-energy when adding more
particles to a fixed-size container).

The way this is reflected in the formalism is that fermions are
described by anticommuting quantities. Formally, we go from real to
complex numbers by doing computations in polynomials of a
newly-introduced variable \(i\), and see what remains if we introduce
the rule \(i^2=-1\). Having completed this step, one finds that these
objects still satisfy the known rules for arithmetics (i.e.~form a
field), but some structure was lost -- there no longer is a notion of
``ordering''. In order to build a suitable mathematical structure for
modeling fermions, one repeats this step, again considering polynomials
over the complex numbers, now formally in more than one variable,
\(\psi_0,\psi_1,\ldots\psi_k\) (in general, one might find a need to
introduce infinitely many), and with a different computational rule:
\(\psi_a\psi_b=-\psi_b\psi_a\), which then in particular implies that
\emph{each such new fundamental variable squares to zero}:
\(\psi_a\psi_a=-\psi_a\psi_a=0\). In basic quantum mechanics, the matrix
element \(U_{23}\) is the coefficient of an operator
\(U_{23}a^\dagger_2a_3\) that removes a particle in state (e.g.~at
position) \(\#3\) and adds a particle in state \(\#2\). The
corresponding operation for fermionic particles must intrinsically
prevent creating two fermions in the same state.

Ultimately, if we want to get any predictions from such calculations, we
are looking for probabilities, which are described by real numbers. So,
this raises the question how any quantity that carries some of these
anti-commuting \(\psi_a\)-factors could ever end up producing a
contribution to a real number? Specifically, if the initial-state and
end-state only contains non-fermions, perhaps two photons each, how
could intermediate (virtual) fermions that get created and destroyed in
the scattering process contribute to the amplitude? The answer here is
that for any such intermediate quantities, quantum mechanics asks us to
integrate over all possible values, and mathematical consistency
constraints make integration over an anticommuting number a formal
procedure that will remove such an anticommuting factor.

\hypertarget{gamma-matrices}{%
\subsection{``Gamma Matrices''}\label{gamma-matrices}}

The important point to keep in mind about this construction is that the
anticommutation properties model what happens if we have more than one
fermion around and swap the role of two of them. A different, but
(subtly) related question is how fermionic quantum states behave under
spatial rotations. These two properties are interlinked due to the
Spin-Statistics Theorem, for which there however is no simple and
intuitive proof. It will be useful to develop some geometric intuition
about spinors. The aspiration of this background section is to be basic
enough to in particular also be accessible to readers who work on
computational infrastructure for sparse matrix/tensor operations.

\hypertarget{geometric-intuition}{%
\subsubsection{Geometric Intuition}\label{geometric-intuition}}

Readers with a computer science background might be aware of the use of
quaternions for representing three-dimensional rotations in a
computationally efficient way, requiring fewer multiplications to
compute products of (relative) rotations than working with \(3\times 3\)
orientation matrices would. One curious property of the quaternionic
representation of rotations is that they can discriminate between a
0-degree rotation and a 360-degree rotation. This is related to the
geometric fact that the rotation group, as a three-parameter manifold,
is not simply connected. One can find closed paths (which one may
visualize as time-dependent rotations) that lead from a given 3d
orientation back to itself which however, as a path, cannot be deformed
to a trivial path that stays at the original orientation. This is
sometimes presented intuitively in the form of the famous ``Dirac belt
trick'': A belt (or, in some demonstrations, a human arm) can be
regarded as modeling a path (parametrized by a
length-walked-along-the-belt coordinate) through the rotation group if
we mark the symmetry axis (running length-wise) of the (flattened out)
belt, introduce orthogonal coordinates ``forward'', ``left hand side
orthogonal and in-plane'', and ``upwards out-of-plane'' at every point
on the symmetry axis. Now, if we put the belt into a configuration where
the associated 3d coordinate systems at both ends of the belt are
orientation-aligned, there are two different and inequivalent cases:
either the belt can be flattened out without any turning at the ends, or
it cannot. In the latter case, rotating one end by 360 degrees gets us
to the former case, so there really are only two different options.
Intuitively, this then means that, when starting from a
coordinate-aligned body-orientation in 3d, there are exactly two
inequivalent (with respect to continuous deformations of a
rotation-sequence) ways to put a body into a given target-orientation
that differ by a 360-degree rotation. While vectors do not notice a
360-degree rotation, and hence neither do any objects described
exclusively with vector language (such as coordinate systems,
ellipsoids, etc.), there actually are mathematical objects on which
rotations are represented as finite-dimensional linear mappings (which
is not the case for belts), but for which a 360-degree rotation is not
the identity: the quaternions witness this claim.

Now, in terms of deformable rotation-sequences, performing two
consecutive 360-degree rotations must always be indistinguishable from
performing no rotation at all -- the only nontrivial cover of the 3d
rotation group is its double cover. So, whatever matrix represents a
360-degree rotation nontrivially must square to an identity matrix.
Also, we notice that a 360-degree rotation always commutes with any
rotation (i.e.~it is a ``central element'' of the rotation group). If
such a matrix were to have eigenvalues +1 and -1, one would find that
there is a choice of basis on the underlying vector space such that the
leading vector-coefficients pick up a factor +1 and the trailing ones a
factor of -1, and coefficients do not mix across these two groups under
any rotation, i.e.~all rotation-matrices would be block-diagonal. One
hence could split vectors into independent parts, for one of which the
360-degree rotation is represented by the minus-identity matrix. At this
point, it may be possible to split vectors further, so two important
questions are whether there is some ``minimal-size'' linear
representation of rotations that can discriminate 0-degree and
360-degree rotations, and whether there is more than one variant of
these. In three spatial dimensions, there is only one such minimal-size
representation, which is equivalent to quaternions, but in other
(spatial or space-time) dimensions, we may have two inequivalent ones.
In eleven-dimensional space-time, there also is only one option.

The higher-dimensional rotation groups (both for spatial as well as
space-time rotations) can be double-covered in just the same way as the
three-dimensional rotation group, and while we in general cannot
introduce numbers that (at least in some form) support all arithmetic
operations, in particular division, as a kind-of higher dimensional
generalization of quaternions, the Gamma matrices serve the equivalent
purpose while not being related to ``numbers'' (except possibly in eight
dimensions). Geometric objects which transform linearly under rotations
(i.e.~we can define rotation matrices on them) in such a way that the
360-degree rotation is represented by a negative-identity are called
``spinorial''. In three dimensions, while one might have thought of
``vectors'' as being a fundamental building block in the sense that
every other geometric object (such as: an oriented area-element) could
be built from them, this is not quite true -- spinors are ``more
fundamental'' in the sense that they can represent geometric objects
that vectors cannot (the ones that can notice a 360-degree rotation),
but one can build vectors (and hence everything else) out of spinors.
This somewhat parallels the observation that from consecutive
application of spatial rotations, one can only obtain other spatial
rotations, but consecutive application of spatial reflections (across
differently-oriented planes) gives rise to \emph{both} reflections and
rotations, so reflections are ``more fundamental''. In general, in
spacetime dimensions where there are two flavors of fundamental spinors,
we will need both to build all geometric objects (or, alternatively,
just one of them plus something else that then cannot be built from them
alone, such as the vectors).

In three dimensions, the quaternionic representation of a rotation by
the angle \(\theta\) around a unit-normal axis \(\vec n\) is:

\begin{equation}R_{\theta,\vec n}=\begin{scriptsize}\left(\begin{array}{rr}+1&0\\0&+1\end{array}\right)\end{scriptsize}\cdot\cos(\theta/2)+\sum_k n_k\sigma_k\sin(\theta/2)\label{eq:quaternionRotation}\end{equation}

where \(\sigma_k\) is the k-th Pauli spin matrix. This is equivalent to
quaternions with imaginary units \(I,J,K\) satisfying
\(I^2=J^2=K^2=IJK=-1\) if one identifies \(I=i\sigma_x\),
\(J=i\sigma_y\), \(K=i\sigma_z\). The rotation matrices, in the above
form, then are complex \(2\times 2\) matrices with determinant +1
(i.e.~elements of \(SU(2)\)), acting on complex 2-dimensional vectors.
They are sequenced by matrix-multiplication, and operate on spinors of
3d space, but in order to rotate vectors, we need to know how vectors
and spinors are related. Obviously, since spinors notice 360-degree
rotations and vectors do not, we should try to obtain vectors by some
linear projection starting from the tensor product of two spinors.
Fortunately, we do have an object readily available that allows us to
connect to two spinor-indices and produce a vector-index, the
\(\sigma_{iAB}\). So, if \(q_A\) is a complex 2-component spinor,
\(q_A^*\sigma^i_{AB}q_B=x^i\) are the components of a real 3-component
vector. A quantum mechanic would want to write this expression which
obtains a vector from a spinor as \(\langle q|\sigma^i q\rangle\), or
alternatively, \({\rm tr}\,\left(\sigma_i|q\rangle\langle q|\right)\).

Three observations about the half-angle formula above are worthwhile:
First, for a small-angle rotation, the correction to first order in the
angle to a given spinor is proportional to the Pauli matrices. In this
sense, they generate small rotations on spinors. Second, the half-angle
makes a 360-degree rotation the minus-identity. Third, for
\(\theta/2=90^\circ\), the sine-term becomes one and the cosine-term is
zero. This shows that we can interpret the three \(\sigma_i\) both as
generators of small rotations on spinors, but \emph{also} as
implementing 180-degree rotations around the coordinate-axes.

One readily convinces oneself that for vectors, two consecutive
180-degree rotations around orthogonal coordinate axes produce the same
orientation, a 180-degree rotation around the third axis. This is
easiest to realize (and it is this view that generalizes to the
interpretation of sigma matrices to higher dimensions in the form of
Gamma matrices) if one views a 180-degree rotation around one axis as an
operation where the coordinate along that axis is kept fixed, and the
coordinates along all perpendicular axes are sign-flipped. Flipping
e.g.~\texttt{(-\/-+)} followed by flipping \texttt{(+-\/-)} produces
\texttt{(-+-)}. So, in 3d, 180-degree rotations on vectors satisfy

\begin{equation}R^{180^\circ}_{x}R^{180^\circ}_{y}=R^{180^\circ}_{y}R^{180^\circ}_{x}\label{eq:r180r180}\end{equation}

This means that we have
\(R^{180^\circ}_{a}R^{180^\circ}_{b}-R^{180^\circ}_{b}R^{180^\circ}_{a}=0\)
for \(a\neq b\).

One can however convince oneself (such by using a belt) that one should
see the two different orderings as leading to results that differ by a
\(360^\circ\)-rotation, which is represented by the identity on vectors,
but the minus-identity on spinors. Hence, we are led to looking for
geometric objects on which rotations are represented by
rotation-matrices, but for which the operations ``sign-flip all
coordinates except one that was singled out'' are represented by
matrices \(\Gamma^i\) satisfying:

\begin{equation}\Gamma^i\Gamma^j = -\Gamma^j\Gamma^i\;\mbox{for}\;i\neq j.\label{eq:CliffordGammaIneqJ}\end{equation}

What about \(i=j\)? Applying two such coordinate-axis-flip operations
consecutively should be an identity, so we would want
\(\Gamma^i\Gamma_i\) to be an identity-matrix \(I\). We hence are
looking for objects with the property: \begin{equation}
\Gamma^i\Gamma^j+\Gamma^j\Gamma^i=2\delta^{ij}\cdot I
\label{eq:CliffordGamma}\end{equation}

One notes that if we have some such set of \(D\) Gamma-matrices in
\(D\)-dimensional space, then tweaking the set by multiplying the first
such gamma matrix with the imaginary unit \(-i\) changes the
\(\delta^{ij}\) on the right hand side to a
\({\rm diag}(-1,+1,\ldots,+1)\) matrix, i.e.~a Minkowski-type scalar
product, while retaining the anticommutation properties. If we start
index-counting with \(0\), then the original ``timelike-coordinate
sign-flip'' becomes \(i\Gamma^0\), and flipping all four
coordinate-axes, i.e. reverting both the time-axis and the
left/right-handedness of space is expressed by
\(i\Gamma^0\Gamma^1\Gamma^2\Gamma^3\) (where each of the four \(\Gamma\)
flips all coordinate axes except one, so every coordinate-axis gets
flipped three times), which is generally also called \(\Gamma^*\), but
for both historical and other geometric reasons in four spacetime
dimensions also \(\Gamma^5\).

\hypertarget{construction}{%
\subsubsection{Construction}\label{construction}}

When a need arises to work with explicit matrix representations of Gamma
matrices (which for many calculations can be avoided!), one can proceed
by a recursive procedure, leveraging the anticommutation properties of
the Pauli spin matrices and general properties of the Kronecker (tensor)
product. We will here gloss over important subtleties that are related
to the fact that the centers (i.e.~elements that commute with all
rotations) of different \(Spin(p, q)\) rotation group look different in
different dimensions, as well as the related algebraic and geometric
perspectives on this fact, which ultimately boils down to the question
whether the Gamma matrices generated by the procedure explained in the
following can be simultaneously block-diagonalized, and also whether
some such blocks can be made all-real. A good discussion of this can be
found in the appendix of the textbook {[}34{]}. Rather, we will (in the
style of the textbook {[}35{]}, which needs to introduce general
\(D\)-dimensional Gamma matrices as a first step towards generalization
to \(D\in\mathbb{C}\)) introduce a simple recursive procedure to obtain
Gamma-matrices for even-dimensional Euclidean space, \(D=2K\). While
there is a better alternative, we could obviously cover the
odd-dimensional case by going to the next-higher even-dimensional one
and remove one Gamma-matrix. There are multiple possible (equivalent)
recursive constructions. The one used here is not commonly presented in
textbooks.

For \(D=2K=2\), it is actually not true that we cannot discriminate a
\(720^\circ\)-rotation from a \(360^\circ\)-rotation. We still want to
look for two matrices that formally satisfy
\(\Gamma^I\Gamma^J+\Gamma^J\Gamma^I=2\delta^{IJ}\) for this recursive
base case. This is easy enough -- we can just take, for example, the
Pauli spin matrices \(\sigma^2=\sigma^y\) and \(\sigma^3=\sigma^z\).

Now, going from \(D=2K\) to \(D+2\), we need to \emph{double} the
dimensionality of our spinors. This is easiest to understand by
realizing that in \(D=2K\), we can find \(K\) coordinate-planes \(P_m\)
spanned by axes \(\vec e_{2m}\) and \(\vec e_{2m+1}\). If we introduce
the operators \(H_m\) that generate rotations in each of the the \(P_m\)
planes, normalized such that a \(2\pi\)-rotation does a full turn, then
the angular momentum for a spinor ``measured'' by this rotation-operator
must have eigenvalues \(\lambda\in\{-1/2,+1/2\}\) (which makes
\(\exp(i\alpha\lambda H_m)=-I\) for \(\alpha=2\pi\)). This is just a
re-statement of the claim that any way to rotate by \(360^\circ\) must
change the sign. So, by induction, in \(D=2K\), we have eigenvalues
\(\pm1/2\) for each of \(K\) independently-measurable angular momenta.
If we encounter angular momentum \(+1/2\), we can ``rotate the system to
turn that plane upside-down'', using other rotation-directions, and so
we also must encounter angular momentum \(-1/2\) along that direction.
If we were to go to a basis of ``definite angular momentum for each
\(H_m\)'', the simultaneous-eigenstates would be labeled with
\((\pm1/2,\pm1/2,\ldots,\pm1/2)\) tags. We can show by construction that
this can be accomplished without any further degeneracy in eigenstates.
One might, however, be able to reduce the total number of eigenstates by
2 or 4, such as by realizing that flipping one angular momentum (such as
that of the \(2-3\)-plane) must always flip an odd number of other
angular momenta (such as that of the \(4-5\) plane when doing the flip
with a 180-degree rotation in the \(3-4\)-plane). It so turns out that
this subtlety is best discussed as an afterthought on the basic
construction.

Forgetting about the simultaneous-angular-momenta-eigenstate space basis
again for a moment, how can we conveniently construct a
\(2^{D/2}\)-dimensional spinor-space from a \(2^{(D-2)/2}\)-dimensional
one? Ultimately, we only want to satisfy the Clifford algebra relations
that each basis Gamma-matrix squares to a unit matrix, and anticommutes
with all others. This is simple enough to achieve: if we form the
Kronecker product of each of the \(D-2\) Gamma matrices we have with
\(\sigma_x\), they still anticommute and square to the identity-matrix
in a double-sized vector space. We then can supplement this with two
more Kronecker products, \(\sigma_y\otimes I_{2^{(D-2)/2}}\) and
\(\sigma_z\otimes I_{2^{(D-2)/2}}\), which anticommute among themselves
due to \(\sigma_y\sigma_z+\sigma_z\sigma_y=0\), and anticommute with
each of the lifted Gamma-matrices due to the identity-matrix commuting
with everything and \(\sigma_x\sigma_{y|z}+\sigma_{y|z}\sigma_x=0\).

This construction is performed by the following Python code snippet:

\begin{Shaded}
\begin{Highlighting}[]
\ImportTok{import}\NormalTok{ functools, itertools, operator, numpy}


\NormalTok{\_SIGMA\_E, \_SIGMA\_X, \_SIGMA\_Y, \_SIGMA\_Z }\OperatorTok{=}\NormalTok{ \_SIGMAS }\OperatorTok{=}\NormalTok{ numpy.asarray(}
\NormalTok{    [[[ }\DecValTok{1}\NormalTok{,  }\DecValTok{0}\NormalTok{ ], [ }\DecValTok{0}\NormalTok{,  }\DecValTok{1}\NormalTok{ ]],}
\NormalTok{     [[ }\DecValTok{0}\NormalTok{,  }\DecValTok{1}\NormalTok{ ], [ }\DecValTok{1}\NormalTok{,  }\DecValTok{0}\NormalTok{ ]],}
\NormalTok{     [[ }\DecValTok{0}\NormalTok{, }\OperatorTok{{-}}\OtherTok{1j}\NormalTok{], [}\OtherTok{1j}\NormalTok{,  }\DecValTok{0}\NormalTok{ ]],}
\NormalTok{     [[ }\DecValTok{1}\NormalTok{,  }\DecValTok{0}\NormalTok{ ], [ }\DecValTok{0}\NormalTok{, }\OperatorTok{{-}}\DecValTok{1}\NormalTok{ ]]])}


\KeywordTok{def}\NormalTok{ generate\_gammas(dim):}
  \ControlFlowTok{if}\NormalTok{ dim }\OperatorTok{\%} \DecValTok{2}\NormalTok{:}
    \ControlFlowTok{raise} \PreprocessorTok{ValueError}\NormalTok{(}\StringTok{\textquotesingle{}Dimension must be even.\textquotesingle{}}\NormalTok{)}
  \ControlFlowTok{if}\NormalTok{ dim }\OperatorTok{==} \DecValTok{2}\NormalTok{:}
    \ControlFlowTok{return}\NormalTok{ \_SIGMAS[}\DecValTok{1}\NormalTok{:}\DecValTok{3}\NormalTok{]  }\CommentTok{\# sigma\_x and sigma\_y.}
\NormalTok{  gammas\_prev }\OperatorTok{=}\NormalTok{ generate\_gammas(dim }\OperatorTok{{-}} \DecValTok{2}\NormalTok{)}
\NormalTok{  dim\_v\_prev, dim\_s\_prev, \_ }\OperatorTok{=}\NormalTok{ gammas\_prev.shape}
\NormalTok{  dim\_s }\OperatorTok{=} \DecValTok{2} \OperatorTok{*}\NormalTok{ dim\_s\_prev}
\NormalTok{  gamma\_star\_prev }\OperatorTok{=}\NormalTok{ numpy.eye(dim\_s\_prev)  }
\NormalTok{  ext\_prev }\OperatorTok{=}\NormalTok{ numpy.einsum(}\StringTok{\textquotesingle{}ab,iAB{-}\textgreater{}iaAbB\textquotesingle{}}\NormalTok{,}
\NormalTok{                          \_SIGMA\_Z,}
\NormalTok{                          gammas\_prev).reshape(}
\NormalTok{                              dim\_v\_prev, dim\_s, dim\_s)}
\NormalTok{  new1 }\OperatorTok{=}\NormalTok{ numpy.einsum(}\StringTok{\textquotesingle{}ab,AB{-}\textgreater{}aAbB\textquotesingle{}}\NormalTok{,}
\NormalTok{                      \_SIGMA\_X,}
\NormalTok{                      gamma\_star\_prev).reshape(dim\_s, dim\_s)}
\NormalTok{  new2 }\OperatorTok{=}\NormalTok{ numpy.einsum(}\StringTok{\textquotesingle{}ab,AB{-}\textgreater{}aAbB\textquotesingle{}}\NormalTok{,}
\NormalTok{                      \_SIGMA\_Y,}
\NormalTok{                      gamma\_star\_prev).reshape(dim\_s, dim\_s)}
  \ControlFlowTok{return}\NormalTok{ numpy.concatenate(}
\NormalTok{      [ext\_prev,}
\NormalTok{       new1[numpy.newaxis, :, :],}
\NormalTok{       new2[numpy.newaxis, :, :]],}
\NormalTok{      axis}\OperatorTok{=}\DecValTok{0}\NormalTok{)}


\KeywordTok{def}\NormalTok{ verify\_clifford\_algebra(gammas):}
\NormalTok{    dim\_v, dim\_s, \_ }\OperatorTok{=}\NormalTok{ gammas.shape}
    \ControlFlowTok{for}\NormalTok{ i, j }\KeywordTok{in}\NormalTok{ itertools.product(}\OperatorTok{*}\NormalTok{[}\BuiltInTok{range}\NormalTok{(dim\_v)]}\OperatorTok{*}\DecValTok{2}\NormalTok{):}
\NormalTok{      g2 }\OperatorTok{=}\NormalTok{ gammas[i] }\OperatorTok{@}\NormalTok{ gammas[j] }\OperatorTok{+}\NormalTok{ gammas[j] }\OperatorTok{@}\NormalTok{ gammas[i]}
\NormalTok{      expected }\OperatorTok{=}\NormalTok{ numpy.allclose(}
\NormalTok{          g2, }\DecValTok{2} \OperatorTok{*}\NormalTok{ numpy.eye(dim\_s) }\OperatorTok{*}\NormalTok{ (i }\OperatorTok{==}\NormalTok{ j))}
      \BuiltInTok{print}\NormalTok{(}\SpecialStringTok{f\textquotesingle{}Gamma[}\SpecialCharTok{\{}\NormalTok{i}\SpecialCharTok{\}}\SpecialStringTok{]Gamma[}\SpecialCharTok{\{}\NormalTok{j}\SpecialCharTok{\}}\SpecialStringTok{] + Gamma[}\SpecialCharTok{\{}\NormalTok{j}\SpecialCharTok{\}}\SpecialStringTok{]Gamma[}\SpecialCharTok{\{}\NormalTok{i}\SpecialCharTok{\}}\SpecialStringTok{]: \textquotesingle{}}
            \SpecialStringTok{f\textquotesingle{}}\SpecialCharTok{\{}\NormalTok{expected}\SpecialCharTok{\}}\SpecialStringTok{\textquotesingle{}}\NormalTok{)}

\NormalTok{test\_gammas }\OperatorTok{=}\NormalTok{ generate\_gammas(}\DecValTok{14}\NormalTok{)}
\NormalTok{verify\_clifford\_algebra(test\_gammas)}
\end{Highlighting}
\end{Shaded}

In general, in any study of relativity that also involves matter
(i.e.~fermions), Gamma matrices will show up, and explicit calculations
will ask for some specific choice. Conventions then depend on what has
been culturally established to handle in particular the specific
dimensionality-reduction that might be possible in the given spacetime
dimension. One particularly nice perk of the construction shown above
(for Euclidean gamma matrices) is that it admits doing explicit
entry-evaluation in one's head(!). This then can of course also be
translated to some very simple code that performs the action of some
particular gamma with an extremely compact bit-twiddling algorithm. This
is conveniently verified with the following helper:

\begin{Shaded}
\begin{Highlighting}[]
\KeywordTok{def}\NormalTok{ show\_bit\_gamma(g):}
\NormalTok{  dim\_s }\OperatorTok{=}\NormalTok{ g.shape[}\DecValTok{0}\NormalTok{]}
\NormalTok{  nbits }\OperatorTok{=}\NormalTok{ dim\_s.bit\_length() }\OperatorTok{{-}} \DecValTok{1}
  \ControlFlowTok{for}\NormalTok{ n1, g\_n1 }\KeywordTok{in} \BuiltInTok{enumerate}\NormalTok{(g):}
\NormalTok{    n2 }\OperatorTok{=}\NormalTok{ numpy.argmax(g\_n1.astype(}\BuiltInTok{bool}\NormalTok{))}
\NormalTok{    c }\OperatorTok{=}\NormalTok{ g\_n1[n2]}
    \BuiltInTok{print}\NormalTok{(}\SpecialStringTok{f\textquotesingle{}g[}\SpecialCharTok{\{}\BuiltInTok{bin}\NormalTok{(dim\_s}\OperatorTok{+}\NormalTok{n1)[}\OperatorTok{{-}}\NormalTok{nbits:]}\SpecialCharTok{\}}\SpecialStringTok{, \textquotesingle{}}
          \SpecialStringTok{f\textquotesingle{}}\SpecialCharTok{\{}\BuiltInTok{bin}\NormalTok{(dim\_s}\OperatorTok{+}\NormalTok{n2)[}\OperatorTok{{-}}\NormalTok{nbits:]}\SpecialCharTok{\}}\SpecialStringTok{]: }\SpecialCharTok{\{}\NormalTok{c}\SpecialCharTok{:+.2f\}}\SpecialStringTok{\textquotesingle{}}\NormalTok{)}
\end{Highlighting}
\end{Shaded}

Splitting spinor indices back into binary multi-indices then provides a
``bit-twiddling'' {[}36{]} perspective onto Gamma matrices.

If we start index counting at zero (as we really should!), we observe
that \(\Gamma^{2k}\) flips the \(k\)-th bit (by increasing order of
significance, i.e.~the \(2^k\)-binary-digit) and also multiplies the
coefficient with -1, but only if there are in total an odd number of
higher-value bits in the index. Correspondingly, the \(\Gamma^{2k+1}\)
also flips the \(k\)-th bit, but also introduces a factor \(i\), and a
further factor \(-1\) if the properties ``number of higher-value bits is
even'' and ``we flip a 0 to a 1'' do \emph{not} align -- so, we get a
factor \(+i\) when the \(k\)-th least significant index-bit is flipped
from 0 to 1 and there is an even number of higher-valued 1-bits (or we
flip from 1 to 0 and there is an odd number), and a factor \(-i\) if we
flip from 1 to 0 and there is an even number of higher-valued 1-bits (or
if we flip from 0 to 1 and there is an odd number of higher-valued
1-bits). With a bit of practice, it then becomes very easy to ``see''
things such as ``applying \(\Gamma^{023}\) to \(\psi_17\) must give
\(-i\psi_{16}\), since we can apply this in order \(\Gamma^{230}\) with
no overall sign change and are flipping the \(2^0\)-bit once and the
\(2^1\)-bit twice. When we flipped the \(2^0\) bit, there was one
higher-valued bit set, giving us a factor \(-1\). When we then apply
\(\Gamma^3\), we flip the \(2^1\)-bit from 0 to 1, there is one
higher-valued bit set, and we are going from 0 to 1. This cancels the
\(-1\) and introduces a factor \(+i\). We then flip the \(2^1\)-bit
again with \(\Gamma^2\), and have one higher-value bit set, so we get
another factor -1. In total, we end up with \(1000_{\rm b}=16\) and an
overall factor \(-i\), so \(\Gamma^{023}\psi_{17}=-i\psi_{16}\). While
this may look like merely a cool party trick, being able to see such
relations quickly can be very helpful when debugging code where it is
unclear if results are off.

While it is certainly nice to be able to think about Gamma matrices in
terms of such simple bit-twiddling operations, the conventions commonly
used in the theoretical physics literature mostly do not care about
admitting such simple approaches. This is unfortunate, but also partly
understandable due to both Lorentz signature and also irreducible
representations sometimes being half-size or quarter-size
(``Majorana/Weyl/Majorana-Weyl'') somewhat complicating the picture. The
problem to come up with appealing conventions for fundamental spinors
across all relevant spacetime dimensions is left as an exercise to the
reader.

Two further notes are perhaps of interest: First, having a set of \(2D\)
Gamma-matrices, one can form nilpotent matrices
\(\Gamma^+_k=\Gamma_{2k}+i\Gamma_{2k+1}\) which satisfy
\(\Gamma^+_i\Gamma^+_j=-\Gamma^+_j\Gamma^+_i\) and square to zero.
Intuitively, these raise the \(k\)-th angular momentum by \(+1\), but
only ``where they can''. This gives us a way to provide a matrix
realization for the anticommuting quantities used to describe fermions
-- if we wanted to interpret them in such a way. Second, the fact that
spinors grow exponentially in dimensionality with increasing spacetime
dimension also puts a brake on the maximal dimensionality that a
physical theory with vector-spinor gauge-particles can have, since the
corresponding dimensionalities for other particles only grows
polynomially with dimensionality, and there can be only one type of
graviton in any variant of General Relativity. After a straightforward
analysis that necessitates some representation theoretic care
(especially due to physical degrees of freedom of massless particles
having transversal polarization), one finds that this maximal dimension
is \emph{eleven}.

\hypertarget{references}{%
\section*{References}\label{references}}
\addcontentsline{toc}{section}{References}

\hypertarget{refs}{}
\begin{CSLReferences}{0}{0}
\leavevmode\vadjust pre{\hypertarget{ref-matzner1967gravitational}{}}%
\CSLLeftMargin{{[}1{]} }
\CSLRightInline{Matzner RA, Misner CW. Gravitational field equations for
sources with axial symmetry and angular momentum. Physical Review
1967;154:1229.}

\leavevmode\vadjust pre{\hypertarget{ref-mars2001spacetime}{}}%
\CSLLeftMargin{{[}2{]} }
\CSLRightInline{Mars M. Spacetime ehlers group: Transformation law for
the weyl tensor. Classical and Quantum Gravity 2001;18:719.}

\leavevmode\vadjust pre{\hypertarget{ref-geroch1972method}{}}%
\CSLLeftMargin{{[}3{]} }
\CSLRightInline{Geroch R. A method for generating new solutions of
einstein's equation. II. Journal of Mathematical Physics
1972;13:394--404.}

\leavevmode\vadjust pre{\hypertarget{ref-breitenlohner1987geroch}{}}%
\CSLLeftMargin{{[}4{]} }
\CSLRightInline{Breitenlohner P, Maison D. On the geroch group. Annales
de l'IHP physique th{é}orique, vol. 46, 1987, p. 215--46.}

\leavevmode\vadjust pre{\hypertarget{ref-belinskii1982general}{}}%
\CSLLeftMargin{{[}5{]} }
\CSLRightInline{Belinskii VA, Khalatnikov IM, Lifshitz EM. A general
solution of the einstein equations with a time singularity. Advances in
Physics 1982;31:639--67.}

\leavevmode\vadjust pre{\hypertarget{ref-damour2003cosmological}{}}%
\CSLLeftMargin{{[}6{]} }
\CSLRightInline{Damour T, Henneaux M, Nicolai H. Cosmological billiards.
Classical and Quantum Gravity 2003;20:R145.}

\leavevmode\vadjust pre{\hypertarget{ref-damour2014quantum}{}}%
\CSLLeftMargin{{[}7{]} }
\CSLRightInline{Damour T, Spindel P. Quantum supersymmetric bianchi IX
cosmology. Physical Review D 2014;90:103509.}

\leavevmode\vadjust pre{\hypertarget{ref-damour2022hidden}{}}%
\CSLLeftMargin{{[}8{]} }
\CSLRightInline{Damour T, Spindel P. Hidden kac-moody structures in the
fermionic sector of five-dimensional supergravity. Physical Review D
2022;105:125006.}

\leavevmode\vadjust pre{\hypertarget{ref-damour2009fermionic}{}}%
\CSLLeftMargin{{[}9{]} }
\CSLRightInline{Damour T, Hillmann C. Fermionic kac-moody billiards and
supergravity. Journal of High Energy Physics 2009;2009:100.}

\leavevmode\vadjust pre{\hypertarget{ref-damour2006hidden}{}}%
\CSLLeftMargin{{[}10{]} }
\CSLRightInline{Damour T, Kleinschmidt A, Nicolai H. Hidden symmetries
and the fermionic sector of eleven-dimensional supergravity. Physics
Letters B 2006;634:319--24.}

\leavevmode\vadjust pre{\hypertarget{ref-de2006extended}{}}%
\CSLLeftMargin{{[}11{]} }
\CSLRightInline{De Buyl S, Henneaux M, Paulot L. Extended E8 invariance
of 11-dimensional supergravity. Journal of High Energy Physics
2006;2006:056.}

\leavevmode\vadjust pre{\hypertarget{ref-freedman2012supergravity}{}}%
\CSLLeftMargin{{[}12{]} }
\CSLRightInline{Freedman DZ, Van Proeyen A. Supergravity. Cambridge
university press; 2012.}

\leavevmode\vadjust pre{\hypertarget{ref-knuth1960imaginary}{}}%
\CSLLeftMargin{{[}13{]} }
\CSLRightInline{Knuth DE. A imaginary number system. Communications of
the ACM 1960;3:245--7.}

\leavevmode\vadjust pre{\hypertarget{ref-nakos2020nearly}{}}%
\CSLLeftMargin{{[}14{]} }
\CSLRightInline{Nakos V. Nearly optimal sparse polynomial
multiplication. IEEE Transactions on Information Theory
2020;66:7231--6.}

\leavevmode\vadjust pre{\hypertarget{ref-vermaseren2000new}{}}%
\CSLLeftMargin{{[}15{]} }
\CSLRightInline{Vermaseren JAM.
\href{https://arxiv.org/abs/math-ph/0010025}{New features of FORM}
2000.}

\leavevmode\vadjust pre{\hypertarget{ref-dean2008mapreduce}{}}%
\CSLLeftMargin{{[}16{]} }
\CSLRightInline{Dean J, Ghemawat S. MapReduce: Simplified data
processing on large clusters. Communications of the ACM
2008;51:107--13.}

\leavevmode\vadjust pre{\hypertarget{ref-harris2020array}{}}%
\CSLLeftMargin{{[}17{]} }
\CSLRightInline{Harris CR, Millman KJ, Walt SJ van der, Gommers R,
Virtanen P, Cournapeau D, et al. Array programming with {NumPy}. Nature
2020;585:357--62. \url{https://doi.org/10.1038/s41586-020-2649-2}.}

\leavevmode\vadjust pre{\hypertarget{ref-fischbacher2005planar}{}}%
\CSLLeftMargin{{[}18{]} }
\CSLRightInline{Fischbacher T, Klose T, Plefka J. Planar plane-wave
matrix theory at the four loop order: Integrability without BMN scaling.
Journal of High Energy Physics 2005;2005:039.}

\leavevmode\vadjust pre{\hypertarget{ref-petkovvsek1996}{}}%
\CSLLeftMargin{{[}19{]} }
\CSLRightInline{Petkovšek M, Wilf HS, Zeilberger D. A=b. 1996.}

\leavevmode\vadjust pre{\hypertarget{ref-goldberg1991every}{}}%
\CSLLeftMargin{{[}20{]} }
\CSLRightInline{Goldberg D. What every computer scientist should know
about floating-point arithmetic. ACM Computing Surveys (CSUR)
1991;23:5--48.}

\leavevmode\vadjust pre{\hypertarget{ref-schroeder2009dram}{}}%
\CSLLeftMargin{{[}21{]} }
\CSLRightInline{Schroeder B, Pinheiro E, Weber W-D. DRAM errors in the
wild: A large-scale field study. ACM SIGMETRICS Performance Evaluation
Review 2009;37:193--204.}

\leavevmode\vadjust pre{\hypertarget{ref-sobczyk2012conformal}{}}%
\CSLLeftMargin{{[}22{]} }
\CSLRightInline{Sobczyk G. Conformal mappings in geometric algebra.
Notices of the AMS 2012;59:264--73.}

\leavevmode\vadjust pre{\hypertarget{ref-hestenes2012clifford}{}}%
\CSLLeftMargin{{[}23{]} }
\CSLRightInline{Hestenes D, Sobczyk G. Clifford algebra to geometric
calculus: A unified language for mathematics and physics. vol. 5.
Springer Science \& Business Media; 2012.}

\leavevmode\vadjust pre{\hypertarget{ref-sommer2013geometric}{}}%
\CSLLeftMargin{{[}24{]} }
\CSLRightInline{Sommer G. Geometric computing with clifford algebras:
Theoretical foundations and applications in computer vision and
robotics. Springer Science \& Business Media; 2013.}

\leavevmode\vadjust pre{\hypertarget{ref-wareham2005applications}{}}%
\CSLLeftMargin{{[}25{]} }
\CSLRightInline{Wareham R, Cameron J, Lasenby J. Applications of
conformal geometric algebra in computer vision and graphics. Computer
algebra and geometric algebra with applications: 6th international
workshop, IWMM 2004, shanghai, china, may 19-21, 2004 and international
workshop, GIAE 2004, xian, china, may 24-28, 2004, revised selected
papers, Springer; 2005, p. 329--49.}

\leavevmode\vadjust pre{\hypertarget{ref-hildenbrand2012foundations}{}}%
\CSLLeftMargin{{[}26{]} }
\CSLRightInline{Hildenbrand D. Foundations of geometric algebra
computing. AIP conference proceedings, vol. 1479, American Institute of
Physics; 2012, p. 27--30.}

\leavevmode\vadjust pre{\hypertarget{ref-bayro2021survey}{}}%
\CSLLeftMargin{{[}27{]} }
\CSLRightInline{Bayro-Corrochano E. A survey on quaternion algebra and
geometric algebra applications in engineering and computer science
1995--2020. IEEE Access 2021;9:104326--55.}

\leavevmode\vadjust pre{\hypertarget{ref-grassmann1862ausdehnungslehre}{}}%
\CSLLeftMargin{{[}28{]} }
\CSLRightInline{Grassmann H. Die ausdehnungslehre. vol. 1. Enslin;
1862.}

\leavevmode\vadjust pre{\hypertarget{ref-hestenes1996grassmann}{}}%
\CSLLeftMargin{{[}29{]} }
\CSLRightInline{Hestenes D. Grassmann's vision. Hermann g{ü}nther
gra{ß}mann (1809--1877): Visionary mathematician, scientist and
neohumanist scholar: Papers from a sesquicentennial conference,
Springer; 1996, p. 243--54.}

\leavevmode\vadjust pre{\hypertarget{ref-freedman1976progress}{}}%
\CSLLeftMargin{{[}30{]} }
\CSLRightInline{Freedman DZ, Nieuwenhuizen P van, Ferrara S. Progress
toward a theory of supergravity. Physical Review D 1976;13:3214.}

\leavevmode\vadjust pre{\hypertarget{ref-deser1976consistent}{}}%
\CSLLeftMargin{{[}31{]} }
\CSLRightInline{Deser SD, Zumino B. Consistent supergravity. Phys Lett B
1976;62:335--7.}

\leavevmode\vadjust pre{\hypertarget{ref-cremmer1978supergravity}{}}%
\CSLLeftMargin{{[}32{]} }
\CSLRightInline{Cremmer E, Julia B, Scherk J. Supergravity in theory in
11 dimensions. Physics Letters B 1978;76:409--12.}

\leavevmode\vadjust pre{\hypertarget{ref-arkani2010simplest}{}}%
\CSLLeftMargin{{[}33{]} }
\CSLRightInline{Arkani-Hamed N, Cachazo F, Kaplan J. What is the
simplest quantum field theory? Journal of High Energy Physics
2010;2010:1--92.}

\leavevmode\vadjust pre{\hypertarget{ref-polchinski2005string}{}}%
\CSLLeftMargin{{[}34{]} }
\CSLRightInline{Polchinski J. String theory. 2005.}

\leavevmode\vadjust pre{\hypertarget{ref-collins1985renormalization}{}}%
\CSLLeftMargin{{[}35{]} }
\CSLRightInline{Collins JC. Renormalization: An introduction to
renormalization, the renormalization group and the operator-product
expansion. Cambridge university press; 1985.}

\leavevmode\vadjust pre{\hypertarget{ref-anderson2005bit}{}}%
\CSLLeftMargin{{[}36{]} }
\CSLRightInline{Anderson SE. Bit twiddling hacks,\\
\hbox{\tt http://graphics.stanford.edu/\~{}seander/bithacks.html};
2005.}

\end{CSLReferences}

\end{document}